\documentclass[conference]{IEEEtran}
\IEEEoverridecommandlockouts
\usepackage{cite}
\usepackage{amsmath,amssymb,amsfonts}
\usepackage{graphicx}
\usepackage{textcomp}
\usepackage{xcolor}
\usepackage{algorithm}
\usepackage{algpseudocode}
\usepackage{subcaption}
\usepackage{caption}
\usepackage{bbm}
\usepackage{pgfplots}
\providecommand{\abs}[1]{\lvert#1\rvert } 

\def\BibTeX{{\rm B\kern-.05em{\sc i\kern-.025em b}\kern-.08em
    T\kern-.1667em\lower.7ex\hbox{E}\kern-.125emX}}
\begin{document}

\title{Proportional Fairness through Dual Connectivity in Heterogeneous Networks}

\author{
	\IEEEauthorblockN{Pradnya Kiri Taksande\IEEEauthorrefmark{1},
		Prasanna Chaporkar\IEEEauthorrefmark{1},
		Pranav Jha\IEEEauthorrefmark{1},
		Abhay Karandikar\IEEEauthorrefmark{1}\IEEEauthorrefmark{2}
		}
	\IEEEauthorblockA{\IEEEauthorrefmark{1}Department of Electrical Engineering,
		Indian Institute Technology Bombay, India 400076\\
		Email: \{pragnyakiri,chaporkar,pranavjha,karandi\}@ee.iitb.ac.in}
	\IEEEauthorblockA{\IEEEauthorrefmark{2}Director, Indian Institute Technology Kanpur, India 208016\\
		Email: karandi@iitk.ac.in
	}
}

\maketitle

\begin{abstract}
Proportional Fair (PF) is a scheduling technique to maintain a balance between maximizing throughput and ensuring fairness to users. Dual Connectivity (DC) technique was introduced by the 3rd Generation Partnership Project (3GPP) to improve the mobility robustness and system capacity in heterogeneous networks. In this paper, we demonstrate the utility of DC in improving proportional fairness in the system. We propose a low complexity centralized PF scheduling scheme for DC and show that it outperforms the standard PF scheduling scheme. Since the problem of dual association of users for maximizing proportional fairness in the system is NP-hard, we propose three heuristic user association schemes for DC. We demonstrate that DC, along with the proposed PF scheme, gives remarkable gains on PF utility over single connectivity and performs almost close to the optimal PF scheme in heterogeneous networks. 
\end{abstract}
\section{Introduction}
Heterogeneous Networks (HetNets) consisting of nodes with varying coverage have been introduced to meet the ever-increasing data traffic demands. At the same time, the evolution of smart mobile devices has led to devices with multiple interfaces. To exploit the availability of multiple interfaces, the 3rd Generation Partnership Project (3GPP) has introduced the technique of Dual Connectivity (DC), in which a Mobile Terminal (MT) is connected to two base stations (BSs) simultaneously \cite{36842}. The data transmission is handled by both BSs with independent scheduling performed at each BS. The extension of DC to multiple connections is known as multi-connectivity, in which an MT is connected to more than two BSs simultaneously. \par
Proportional Fair (PF) is a popular scheduling algorithm, which seeks to balance throughput maximization with fairness in the system. PF maximizes the sum of the logarithm of per-user throughput \cite{kushner}, known as PF utility. %In \cite{exploiting}, the authors determine an optimal user association algorithm that maximizes the weighted sum rate system utility as well as and proportional fairness (PF) system utility subject to per-user rate constraints.
%In \cite{alexandris}, the authors propose a utility proportional fairness resource allocation that supports MT requested rates. The authors in \cite{matching} formulate the PF utility maximization problem in LTE-WLAN Aggregation scenario and propose a joint cell selection and power control algorithm.
%In \cite{GenPF}, the authors propose an MT association scheme for balancing the load in the network and improving the PF in a 3G data network. In \cite{MimoPF}, the general PF scheduling scheme is extended to Multiple Input Multiple Output (MIMO) case. The extension of PF to multicarrier systems is shown in \cite{PFmulticarrier}.
The authors in \cite{GPF} introduce a Global PF (GPF) scheduling scheme and prove that GPF maximizes the PF utility when all MTs are connected to all BSs. The basic idea of this scheme is to connect all MTs to every BS and then schedule an MT opportunistically at the BS providing the best channel to it. This distributed scheme, however, requires the sharing of per-user throughput information between each pair of BSs, which leads to significant control signaling exchange. Moreover, the maintenance of multiple connections requires additional resources, for instance, power, radio, and processing resources at the network. As a result, the overhead to maintain multiple connections for a single MT may be significantly high. Due to these reasons, connecting all MTs to every BS is not practical. Further, only a small number of interfaces, typically one or two, are used by an MT. \par
One of the key challenges in mobile networks lies in the selection of BS for MT association. The problem of single or dual association of MTs to maximize the PF utility in the system is known to be NP-hard \cite{GenPF}, \cite{PoA}. Hence, there does not exist a polynomial-time algorithm that can determine the optimal solution. \par
In this paper, we consider a viable scenario where each MT has two interfaces and, therefore, can connect to a maximum of two BSs at the same time. Earlier works \cite{matching,analRealistic} consider two connections for MTs. However, in contrast to \cite{GPF,matching,analRealistic}, we consider a network architecture with a centralized controller \cite{Pradnya} communicating with all BSs. 
%Contributions
Our main contributions are as follows. 
\begin{itemize}
	\item We propose a low complexity centralized PF scheduling scheme for DC (PF-DC). 
	\item Since the PF-DC scheme is centralized and the number of connections for an MT are limited to two, the control signaling overhead in the PF-DC scheme is reduced significantly as compared to that in the GPF scheme \cite{GPF}. We also demonstrate that the PF-DC scheme outperforms the standard PF scheduling scheme for DC in terms of PF utility and average throughput of MTs.
	\item Since the problem of dual association to maximize PF utility is NP-hard, we propose three heuristic association algorithms for DC.
	\item We evaluate the performance of PF-DC in conjunction with these algorithms and compare them with single connectivity and all connectivity. We demonstrate that DC achieves a significant increase in PF utility as compared to single connectivity, and it is close to that provided by all connectivity with GPF scheduling (optimal).
	\item The proposed schemes can be easily employed in real deployment due to reduced overheads and simplicity.
\end{itemize}
The paper is organized as follows. Section \ref{sysModel} describes the system under consideration. The proposed scheduling scheme, PF-DC, is explained in Section \ref{pf-dc}. The complexity analysis of the PF-DC scheme is detailed in Section \ref{complexity}. The MT association algorithms are proposed in Section \ref{propAlgo}. The simulation results are presented in Section \ref{sim}, followed by the conclusion in Section \ref{conc}.
\section{System Description}
\label{sysModel}
\subsection{System Model}
We consider a Software-Defined Networking (SDN) based Radio Access Network (RAN) architecture, as proposed in our previous work \cite{Pradnya}. The control plane functionality is handled by SDN based RAN Controller (SRC), and the data plane is managed individually at each BS. The SRC supports radio resource control and management functions such as admission control, mobility management, and load balancing. Access to the global view of the system enables SRC to take appropriate decisions such as MT association. Consider a RAN consisting of $B$ BSs and $U$ MTs, as illustrated in Figure \ref{scenario}. Let $\mathcal{B}$ and $\mathcal{U}$ denote the set of BSs and MTs, respectively. Let the total number of BSs and MTs be denoted by $\abs{\mathcal{B}}=N$ and $\abs{\mathcal{U}}=M$, respectively, where $\abs{\mathcal{X}}$ denotes the cardinality of set $\mathcal{X}$. The BSs can be macro or pico. The MTs do not differentiate between different types of BSs and can connect to any two of them simultaneously. %Each BS is assigned a fixed bandwidth and transmits at a fixed power. The system functions in a time-slotted manner with fixed duration time-slot.
We assume an infinitely backlogged traffic model for MTs.
\par
\subsection{Standard PF Scheduling}
The system functions in a time-slotted manner with a time-slot of fixed duration. PF scheduling is performed individually at all the BSs. Each BS takes scheduling decisions at the beginning of a time slot. %At each BS, the PF metric for all MTs is determined based on the instantaneous rate that can be received by that MT, and the average throughput which the MT has achieved until that time. For instance, 
The PF metric for MT $j$ at BS $k$ is determined by $ {r^k_j}(t)/{\bar{r}^k_j}(t)$, where ${r^k_j}(t)$ is the achievable rate by MT $j$ at time $t$, and ${\bar{r}^k_j}(t)$ represents the average throughput received by MT $j$ until time $t$. The MT with the maximum PF metric ${u^{\ast}_k} = \arg\max\limits_{j \in \mathcal{U}} {r^k_j}(t)/{\bar{r}^k_j}(t)$ is then selected to be scheduled in slot $t$ by BS $k$. In each time-slot, the average throughput of MT $j$ is updated using a weighted moving average, 
\begin{equation}
\label{pf}
{\bar{r}^k_j}(t+1) = (1-\gamma) \: {\bar{r}^k_j}(t) + \gamma \: {r^k_j}(t) \: \mathbbm{1}{ \{j=u^{\ast}_k \}},
\end{equation} 
where, $\mathbbm{1}\{A\}$ denotes an indicator function of the event $A$, and $0 < \gamma < 1$ is a constant, typically, set to 0.01.
\section{PF-DC: Proposed PF scheduling scheme for dual connectivity}
\label{pf-dc}
In this section, we present PF-DC, the proposed PF scheduling scheme for dual connected MTs. This scheme is similar to the standard PF scheduling scheme with two modifications. First, the PF metric is determined using the sum of the average throughput received by the MT from the two BSs to which it is connected. Each BS $k$ chooses the MT with maximum PF metric (${u^{\ast}_k}$) to schedule in each time-slot as,
\begin{equation}
{u^{\ast}_k} = \arg\max\limits_{j \in \mathcal{U}} \frac {{r^k_j}(t)} {\sum_{k \in \mathcal{B}}{\bar{r}^k_j}(t)}. 
\end{equation}
The denominator is determined by the summation of the average throughput at the two BSs to which the MT is connected. Second, the calculation of the total average throughput of all MTs is performed at the centralized controller (SRC).\par
BSs share the average throughput information regarding their respective dual connected MTs with the SRC. SRC determines the total average throughput of each MT using the shared information and sends it back to the respective BSs. This throughput information is used individually by the BSs for scheduling. The exchange of throughput information between the SRC and BSs takes place at regular intervals with period $T$. We have performed simulations for different values of $T$ (not included here due to space constraints). As $T$ increases, PF utility decreases; the variation in PF utility, however, is not significant. Thus, the value of period $T$ can be tuned as per the system requirements.
%The parameter $T$ can be tuned as per the requirements of the service provider. %Due to the periodic sharing of information, no synchronization is required between BSs and SRC.
Even if there is a slight delay in the sharing of throughput information between SRC and a BS, it affects the scheduling at that BS only. This is because, after receiving information from SRC, scheduling is performed independently at individual BSs. Thus, exact synchronization is not required between BSs and SRC in this scheme.
\begin{figure}%[!htb]
	\centering
	\includegraphics[width=7cm]{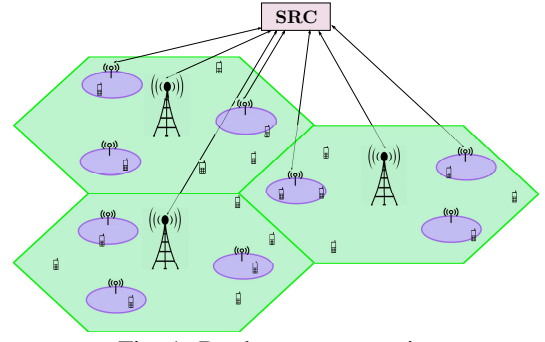}
	\caption{Deployment scenario.}
	\label{scenario}
\end{figure}
\section{Complexity Analysis of PF-DC}
\label{complexity}
In this section, we compare the communication and computation complexity of the PF-DC scheme with that of the GPF scheme \cite{GPF}. We consider only signaling information exchange for our analysis. In the case of GPF, every $T$ time slots, the information is being exchanged between each pair of BSs for every MT. The communication complexity for GPF is $\frac{1}{T} M N (N-1)$. In our network architecture, SRC has access to all BSs in the RAN. Each MT is connected to two BSs, and each BS shares the average throughput information of MTs associated with it with SRC. SRC processes this information and sends the total throughput information back to the BSs every $T$ time slots. Hence, the communication complexity is $\frac{1}{T}\times 2\times2\times M$.\par
The basic computations required by the PF scheduling algorithm are comparisons, additions, and multiplications. For choosing the MT to be scheduled in a scheduling interval, PF scheduling determines the MT with the maximum PF metric. For this, GPF requires $M-1$ comparisons per BS, i.e., a total of $N(M-1)$ comparisons. In the case of PF-DC, let each BS $i$ has $y_i$ connections. Then the total number of comparisons is $\sum_{i=1}^{N} (y_i-1)$. This comes out to be $\sum_{i=1}^{N} y_i - N$. The total number of connections for $M$ MTs is $2M$ since we allow each MT to have two connections. Thus, the number of comparisons is $2M-N$. The number of additions and multiplications for GPF are $NM+\frac{1}{T}(N-1)M$ and $3NM$, respectively, whereas, for PF-DC, this comes out to be $2M+\frac{1}{T}M$ and $6M$, respectively. Thus, we see that the communication and computation complexity of PF-DC is significantly less as compared to the GPF scheme. \par
\section{MT Association Algorithms for Dual Connectivity}
\label{propAlgo}
The problem of association of dual connected MTs for maximizing proportional fairness in the system is NP-hard \cite{PoA}. %Moreover, the solution proposed in \cite{PoA} requires the knowledge of the achievable rate of MTs to every BS, which is not readily available in practice. 
Therefore, we propose heuristic association algorithms for DC in this section. The algorithms are based on Reference Signal Received Power (RSRP) of MTs.
\subsection{User Initiated Greedy with Offloading (UIGO) scheme}
As the name suggests, this scheme is initiated at the MT. Each MT selects two distinct BSs based on RSRP from BSs (see Algorithm \ref{alg2}). $RSRP(a,b)$ represents the RSRP received by MT $b$ from BS $a$. $A_1(b)$ and $A_2(b)$ denote two BSs to which MT $b$ is associated. Each MT selects the BS providing it with maximum RSRP for the first association (Line 2,3). From among the remaining BSs, the MT determines the second BS with maximum RSRP (Line 6). This information is then shared with SRC. For the second association, SRC checks if the second BS is macro (Line 8). MT then determines the third BS with maximum RSRP from among the remaining BSs (Line 9). SRC acquires the third BS details from the MT and checks if it is a small cell. $\mathcal{B}_M$ denotes the set of macro BSs (Line 9). If the difference between RSRP of the second and third BS is within a threshold $H_1$ (Line 10), then SRC selects the third BS, else it selects the second BS. This algorithm enables offloading of MTs from macro cells to small cells if signal strength from a small cell is comparable to that of the macro cell. A global view of the network allows SRC to take appropriate decisions for the second association of MTs. Let $M,N$ denote the number of MTs and BSs, respectively. The computation complexity of this algorithm is $O(MN)$.\par
\begin{algorithm}[!htb]
	\caption{User Initiated Greedy with Offloading (UIGO)}
	\label{alg2}
	\begin{algorithmic}[1]
		\ForAll {$u \in \mathcal{U}$}
		\State $b^{\ast}\gets \arg\max\limits_{b \in \mathcal{B}} (RSRP(b,u))$ 
		\State $A_1(u) \gets b^{\ast}$
		%\State $RSRP(b^{\ast},u) \gets -\infty$
		\EndFor
		\ForAll {$u \in \mathcal{U}$}
		\State $c^{\ast}\gets \arg\max\limits_{b \in \mathcal{B}\setminus \{A_1(u)\}} (RSRP(b,u))$ 
		\State $A_2(u)\gets c^{\ast}$ 
		\If	{$c^{\ast}$ is a macro BS} 
		%\State $RSRP(c^{\ast},u) \gets -\infty$
		\State $d^{\ast}\gets \arg\max\limits_{b \in \mathcal{B} \setminus \mathcal{B}_M } (RSRP(b,u))$ 
		\If	{$(RSRP(c^{\ast},u) - RSRP(d^{\ast},u)) < H_1$} 
		\State $A_2(u)\gets d^{\ast}$ 
		\EndIf
		\EndIf
		\EndFor
	\end{algorithmic}
\end{algorithm}
\subsection{BS Initiated Greedy with User feedback (BIGU) scheme}
This scheme is based on a team selection process where a BS selects a team of MTs. The BSs are sequenced in a round-robin fashion, and each BS chooses an MT in each round (see Algorithm \ref{alg3}). Each MT can be selected twice by two distinct BSs, as we allow two connections per MT. The selection of an MT is based on the RSRP at the MT from the corresponding BS. After the selection of an MT ($u^{\ast}$) by a BS ($b$) (Line 4), the MT checks if the RSRP from this BS is within a threshold ($H_2$) as compared to the RSRP from its best BS ($c^{\ast}$) (Lines 6-8). If it is, MT $u^{\ast}$ is associated with BS $b$. Else, it rejects the current BS $b$ and waits for a better offer from another BS from which it can obtain a higher signal strength. The process continues until two distinct BSs are selected for each MT. SRC coordinates between BSs and MTs to select the appropriate BS for MTs incorporating load balancing in the system. The computation complexity of this algorithm is $O(M^2N)$.\par
\begin{algorithm}[!htb]
	\caption{BS Initiated Greedy with User feedback (BIGU)}
	\label{alg3}
	\begin{algorithmic}[1]
		\State $U_{state}(\cdot) \gets 2$.
		\While{$U_{state}(u) > 0$ for all $u \in \mathcal{U}$ }
		\ForAll {$b \in \mathcal{B}$}
		\State $u^{\ast}\gets \arg\max\limits_{j \in \mathcal{U}} (RSRP(b,j))$ 
		%\If {$U_{state}(u^{\ast}) > 0$}
		\ForAll {$c \in \mathcal{B}$}
		\State $c^{\ast}\gets \arg\max\limits_{c \in \mathcal{B}} (RSRP(c,u^{\ast}))$ 
		\EndFor
		\If	{$RSRP(c^{\ast},u^{\ast}) - RSRP(b,u^{\ast}) < H_2$} 
		\State $U_{state}(u^{\ast}) \gets U_{state}(u^{\ast})-1$
		\If {$U_{state}(u^{\ast}) = 2$}
		\State $A_1(u^{\ast})\gets b$ 
		\ElsIf {$ U_{state}(u^{\ast}) = 1$}
		\State $A_2(u^{\ast})\gets b$ 
		\EndIf
		\State $RSRP(b,u^{\ast}) \gets -\infty$
		%\EndIf
		\EndIf
		\EndFor
		\EndWhile
	\end{algorithmic}
\end{algorithm}
\subsection{Stable Matching (SM) scheme}
The Stable Matching (SM) scheme is a centralized scheme running at the SRC. In this scheme, the set of MTs is ranked by each BS based on their RSRP. Similarly, all BSs are ranked by the MTs based on their RSRP from respective BSs. These preferences are aggregated at the SRC. The size of the sets of BSs and MTs is made equal by repeating each BS $q=M/N+c$ times, where $c$ is a constant. The constant $c$ is selected such that the load for all BSs is almost equal, thus balancing the load in the system. This problem then converts to a stable matching problem, which can be solved using the Gale-Shapley algorithm \cite{galeshapley} (See Algorithm \ref{alg4}). After the first association, the MT preferences are updated by giving the BS of the first association last preference. Similarly, BS preferences are updated. To determine the second association, the problem with updated preferences is again solved using the Gale-Shapley algorithm. The computation complexity of this algorithm is $O(M^2)$. 
\begin{algorithm}[!htb]
	\caption{Stable Matching (SM)}
	\label{alg4}
	\begin{algorithmic}[1]
		\State $P_1 \gets$ Each MT $u$ sets its preferences for all BSs in $\mathcal{B}$
		\State $P_2 \gets$ Each BS $b$ sets its preferences for all MTs in $\mathcal{U}$
		\State Repeat each BS $q=M/N+c$ times for load balancing
		\While {all MTs not allocated to some BS}
		\State $A_1(u) \gets$ Solve stable matching problem using Gale-Shapley algorithm ($P_1,P_2$)
		\EndWhile
		\State $P_1 \gets$ MT $u$ sets its preferences for all BSs in $\mathcal{B}$ giving $A_1(u)$ as least preference
		\State $P_2 \gets$ BS $b$ sets its preferences for all MTs in $\mathcal{U}$
		giving all MTs already associated to it as least preference
		\While {all MTs not allocated to some BS}
		\State $A_2(u) \gets$ Solve stable matching problem using Gale-Shapley algorithm ($P_1,P_2$)
		\EndWhile
	\end{algorithmic}
\end{algorithm}
\section{Numerical Results and Analysis}
\label{sim}
We consider a two-tier HetNet scenario with three macro cells and three pico cells deployed in each macro cell. The operating frequencies of macro and pico BSs are different. The pico cells are deployed using two deployment scenarios. (i) The locations of pico cells are fixed, and their coverages are non-overlapping (e.g., see Figure \ref{scenario}). (ii) Pico cells are randomly deployed within the macro cell coverage and may overlap with other pico cells. The MTs are dropped according to two different deployment scenarios. (i) Hotspot deployment: Two-third MTs are deployed uniformly in the coverage area of pico cells, and one-third MTs are deployed uniformly outside the coverage of pico cells but within the macro cell. (ii) Uniform deployment: MTs are deployed uniformly in the coverage area of the macro cell without any consideration of the locations of pico cells. \par
Using the combination of pico cell deployments and MT deployments, we have four different deployment scenarios as follows: \textbf{Scenario 1:} Pico cells are deployed at fixed locations with non-overlapping coverage, and MTs are dropped using hotspot deployment. \textbf{Scenario 2:} Pico cells are deployed at fixed locations with non-overlapping coverage, and MTs are dropped using uniform deployment. \textbf{Scenario 3:} Pico cells are deployed randomly within macro cells, and MTs are dropped using hotspot deployment. \textbf{Scenario 4:} Pico cells are deployed randomly within macro cells, and MTs are dropped using uniform deployment. \par
We denote the procedure of all MTs connected to every BS and GPF scheduling \cite{GPF} performed at each BS as the All Connectivity Procedure (ACP). The scheme of an MT connected to the BS offering it with the best signal strength and standard PF scheduling performed at each BS is denoted as Single Connectivity Procedure (SCP). The procedure where an MT is dual connected to two suitable BSs using proposed association algorithms and standard PF scheduling performed independently at each BS is denoted as Dual Connectivity Standard Procedure (DCSP). The scheme where an MT is dual connected to two suitable BSs using proposed association algorithms and PF-DC scheduling performed at each BS in conjunction with SRC is denoted as Dual Connectivity Procedure (DCP). \par
\begin{table}[!htb]
	\caption{Network parameters.}
	\begin{center}
		\begin{tabular}{|l|l|}
			\hline
			\textbf{Parameter} &	\textbf{Value (Macro, Pico)} \\
			\hline
			Macro ISD, Pico radius & 500m, 80m\\
			\hline
			Transmit power & 46 dBm, 30 dBm \\
			\hline
			Antenna & Omnidirectional, Omnidirectional \\
			\hline
			Bandwidth & 5MHz, 5MHz \\
			\hline
			Antenna height & 32m, 10m \\
			\hline
			Path loss (d in km) & 128.1 + 37.6 log(d), 
			140.7 + 36.7 log(d) dB\\
			\hline
		\end{tabular}
	\end{center}
	\label{Parameters}
\end{table}
The simulations are performed using the LENA module in ns-3 simulator \cite{NS3}. Path loss and fading are considered in the simulation scenario. The characteristics of the network simulated are enumerated in Table \ref{Parameters}. The simulations are performed for a duration of 10000 time slots, with simulation run repeated 10 times using independent random numbers. \par 
\subsection{Comparison of PF-DC with Standard PF Scheduling}
We compare the DCP and DCSP procedures, i.e., PF-DC and standard PF scheduling schemes for DC, in this section. The DC algorithms proposed in Section \ref{propAlgo} are used for MT association in these procedures. We use PF utility and average MT throughput as system metrics to compare these two procedures. PF utility is a metric that represents a balance between total throughput and fairness in the system. PF utility is given by $\sum_{u \in \mathcal{U}} \log{(x_u)}$, where $x_u$ denotes the average throughput of MT $u$. A high value of PF utility indicates a balance between total throughput and fairness.\par

\begin{figure} [!htb]%[H]
	\centering
	\includegraphics[width=7cm]{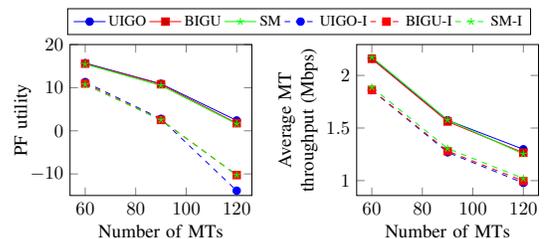}
	\caption{Comparison of PF-DC and standard PF scheduling for DC in Scenario 3.}
	\label{IndPF1}
\end{figure}

\begin{figure} [!htb]%[H]
	\centering
	\includegraphics[width=7cm]{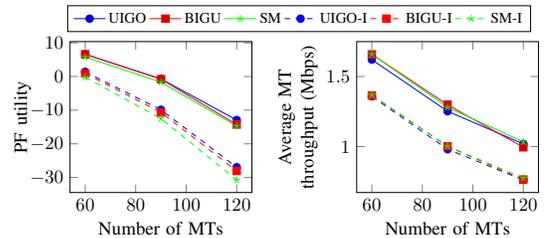}
	\caption{Comparison of PF-DC and standard PF scheduling for DC in Scenario 4.}
	\label{IndPF2}
\end{figure}

Figures \ref{IndPF1} and \ref{IndPF2} illustrate the PF utility and average per-MT throughput metrics in the case of Scenarios 3 and 4, respectively. The plots UIGO, BIGU, SM, represent the MT association chosen according to UIGO, BIGU, SM algorithms, respectively, and the PF-DC scheme used for scheduling. The plots UIGO-I, BIGU-I, SM-I, represent the MT association chosen according to UIGO, BIGU, SM algorithms, respectively, and standard PF scheduling performed independently at each BS. As the number of MTs in the system increases, the average throughput per-MT decreases as the total available capacity is divided among more MTs. We observe a slight improvement in the total throughput with an increase in the number of MTs in the system. However, MTs in different regions within the coverage area of cells receive a varying amount of throughput, and fairness in the system drops. Therefore, there is a drop in the PF utility of the system. Similar results are observed for Scenarios 1 and 2 as well but are omitted here due to space constraints. Thus, PF-DC outperforms the standard PF scheduling scheme not only in terms of the PF utility but in terms of the average MT throughput as well.
\subsection{Comparison of DCP with ACP and SCP}
In this section, we compare the procedures DCP, ACP, and SCP. In DCP, the algorithms proposed in Section \ref{propAlgo}, viz., UIGO, BIGU, and SM, are used for MT association, and the PF-DC scheme is used for scheduling. The period of information exchange $T$ of the PF-DC scheme is considered to be 25 slots. The performance metrics used for comparison are PF utility, system throughput, and Jain's Fairness Index (JFI). JFI denotes the fairness in the throughput values obtained by all MTs in the system. It is given by $\{\sum_{u=1}^{M} x_u\}^2 / M \sum_{u=1}^{M} {x_u}^2 $, where $x_u$ denotes the long term average throughput of MT $u$. The value of JFI closer to one implies a high level of fairness between the MTs. \par
\subsubsection{Scenario 1}
\begin{figure*} [!h]
	\centering
	\includegraphics{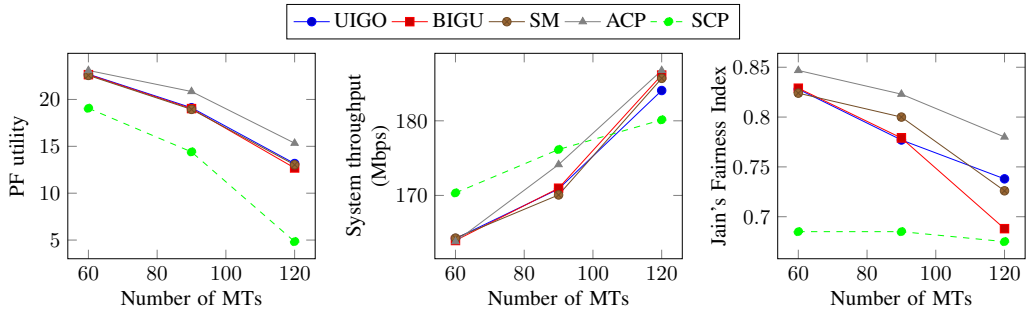}
	\caption{Comparison of various algorithms in Scenario 1.}
	\label{fp1}
\end{figure*}

Figure \ref{fp1} illustrates the variation in PF utility, system throughput, and JFI for various algorithms as a function of the number of MTs in the system in Scenario 1. As the number of MTs in the system increases, system throughput increases due to an increase in multi-user diversity in the system. The per MT throughput in the system, however, drops. At the same time, JFI in the system decreases since variation in received throughput increases with an increase in the number of MTs. Hence, the PF utility decreases with an increase in the number of MTs. We observe that ACP provides high PF utility, i.e., it provides a balance between total throughput and fairness in the system. SCP provides the least PF utility since it neither utilizes the entire capacity in the network nor gives fairness to MTs. The DC association algorithms (UIGO, BIGU, and SM), in conjunction with the PF-DC scheduling algorithm (DCP), provide a substantial improvement in PF utility as compared to SCP, but it is slightly less than ACP. Though ACP provides maximum JFI, the system throughput values for DCP and ACP are almost comparable. \par
\subsubsection{Scenario 2}
\begin{figure*} [!h]
	\centering
	\includegraphics{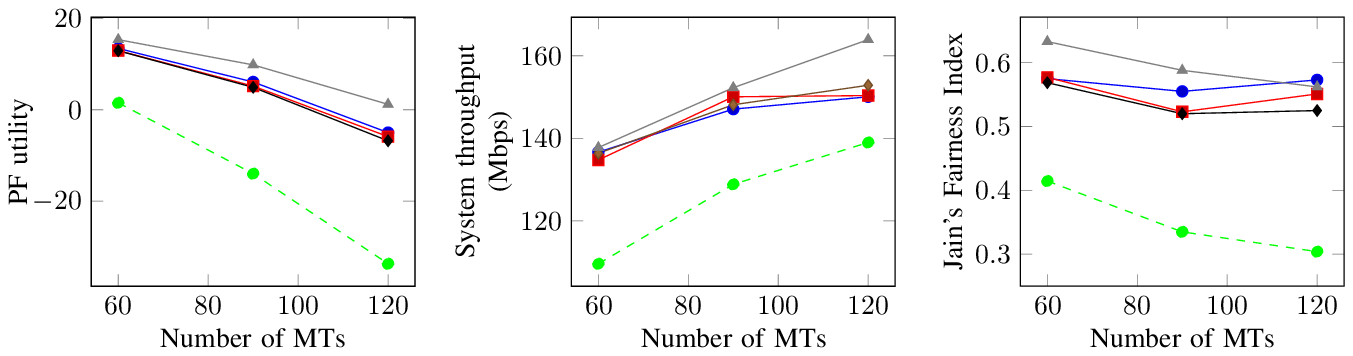}
	\caption{Comparison of various algorithms in Scenario 2.}
	\label{fp2}
\end{figure*}

Figure \ref{fp2} demonstrates the variation in PF utility, system throughput, and JFI for various algorithms as a function of the number of MTs in the system in Scenario 2. In Scenario 2, the MTs may not be deployed inside pico coverage. Hence, SCP is performing worse as compared to its performance in Scenario 1. In this case, SCP gives minimum system throughput and minimum PF utility. ACP provides maximum PF utility as it gives maximum JFI as well as maximum total throughput. Since MTs are uniformly deployed, ACP has an advantage as it balances the load among the BSs by opportunistically choosing the MT with peak rates at each BS. DC algorithms perform slightly worse than ACP due to the limitation in the number of connections, but the total throughput and JFI values are almost comparable for DCP and ACP. 
\subsubsection{Scenario 3}
\begin{figure*} [!h]
	\centering
	\includegraphics{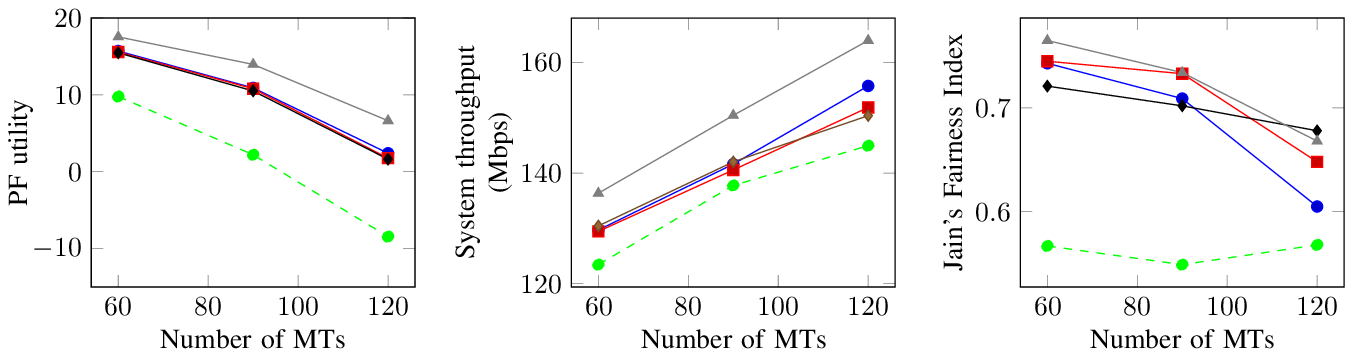}
	\caption{Comparison of various algorithms in Scenario 3.}
	\label{rp1}
\end{figure*}

Figure \ref{rp1} illustrates the variation in PF utility, system throughput, and JFI for various algorithms as a function of the number of MTs in the system in Scenario 3. ACP gives maximum PF utility along with maximum system throughput. The benefit of multiple connections is accentuated by the overlapping coverage of pico cells as well as hotspot deployment of MTs in this scenario. Overlapping coverage works as an advantage for ACP and DCP. However, in SCP, MT connects to the BS from which it receives maximum RSRP. Consider an MT situated in the overlapping coverage area of two cells. In SCP, the MT is connected to one of these two BSs, whereas in DCP and ACP, the MT utilizes the resources of both cells. In SCP, the load among the BSs may not be distributed, and there may be overloaded cells. Hence, as compared to DCP and ACP, SCP gives the worst PF utility. In DCP, a substantial increase in PF utility is observed as compared to that of SCP.
\subsubsection{Scenario 4}
\begin{figure*} [!h]
	\centering
	\includegraphics{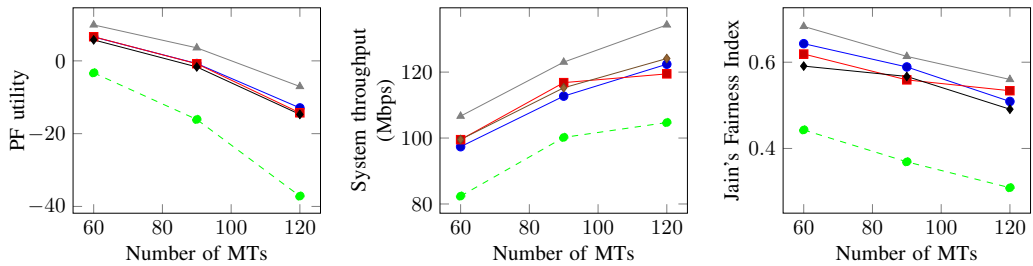}
	\caption{Comparison of various algorithms in Scenario 4.}
	\label{rp2}
\end{figure*}

Figure \ref{rp2} demonstrates the variation in PF utility, system throughput, and JFI for various algorithms as a function of the number of MTs in the system in Scenario 4. ACP maintains maximum PF utility by providing high system throughput and high JFI. The multiple connections offer benefits to the MTs in overlapping pico coverage areas. In SCP, MTs are opportunistically connected to their best BS, thus leading to an imbalance in the system. Thus, SCP provides the lowest PF utility due to weak JFI as well as low total throughput. DC algorithms provide PF utility lower than ACP but much higher than SCP by creating a balance between total throughput and fairness among MTs. \par
We present some general inferences from the results obtained. The values for system throughput and PF utility are higher in general for the hotspot deployment of MTs since more MTs are located in areas where the deployed BSs provide coverage. The performance of proposed association algorithms does not exhibit much variation in terms of the PF utility system metric. This implies that the association of dual connected MTs does not play a major role in the PF utility of the system, but it matters how PF scheduling is performed. There is a remarkable improvement from SCP to DCP in all four scenarios, but the improvement from DCP to ACP is not as much. This improvement from DCP to ACP is even less in Scenario 2. In this scenario, the advantage of multiple connections is available for only a few MTs, and hence, the performance of ACP and DCP is almost comparable. However, as demonstrated in Section \ref{complexity}, ACP incurs an extra computational and communication cost as compared to DCP. 
%
%
%% T = 25msec

%
% T = 25msec
\section{Conclusion}
\label{conc}
We propose a low complexity centralized PF scheduling scheme for DC (PF-DC), which outperforms the standard PF scheduling scheme for DC. The analysis in this paper suggests that the addition of the second radio link (DCP) along with the PF-DC scheduling algorithm brings diversity and bestows remarkable gains in PF utility over SCP. It also demonstrates that further gain in PF utility with additional radio links (ACP) is marginal and comes at a significant additional cost of maintenance of a large number of connections for an MT. Thus, dual connectivity, in conjunction with PF-DC scheduling, improves the proportional fairness in the system. We propose various association algorithms for DC and observe that they give similar performance as far as the PF utility is concerned. Even though we propose to use a centralized RAN controller for the execution of the PF scheduling algorithm, the proposed scheme can easily be employed in existing HetNets as well where one of the BSs can act as the centralized controller for a subset of MTs.
\section{Acknowledgement}
This work has been supported by the Department of Telecommunications, Ministry of Communications, India as part of the Indigenous 5G Test Bed project.
\bibliographystyle{ieeetr}
\bibliography{IEEEabrv,Algo_References}

\begin{thebibliography}{10}

\bibitem{36842}
3GPP, ``{Study on small cell enhancements for E-UTRA and E-UTRAN : Higher layer
  aspects},'' TR 36.842, {3rd Generation Partnership Project}, Jan 2014.
\newblock Version 12.0.0.

\bibitem{kushner}
H.~J. Kushner and P.~A. Whiting, ``Asymptotic properties of proportional-fair
  sharing algorithms,'' in {\em Annual Allerton Conference on Communication,
  Control, and Computing}, pp.~1--6, 2002.

\bibitem{GPF}
H.~Zhou, P.~Fan, and J.~Li, ``Global proportional fair scheduling for networks
  with multiple base stations,'' {\em IEEE Transactions on Vehicular
  Technology}, vol.~60, no.~4, pp.~1867--1879, 2011.

\bibitem{GenPF}
T.~Bu, L.~Li, and R.~Ramjee, ``Generalized proportional fair scheduling in
  third generation wireless data networks,'' in {\em IEEE International
  Conference on Computer Communications}, pp.~1--12, 2006.

\bibitem{PoA}
J.~Lee and S.~Bahk, ``Point of attachment selection in heterogeneous radio
  access technology environments,'' in {\em IEEE Symposium on Computers and
  Communications}, pp.~873--878, 2010.

\bibitem{matching}
Q.~Han, B.~Yang, C.~Chen, and X.~Guan, ``Matching-based cell selection for
  proportional fair throughput boosting via dual-connectivity,'' in {\em {IEEE
  Wireless Communications and Networking Conference (WCNC)}}, pp.~1--6, 2017.

\bibitem{analRealistic}
G.~Pocovi, S.~Barcos, H.~Wang, K.~I. Pedersen, and C.~Rosa, ``Analysis of
  heterogeneous networks with dual connectivity in a realistic urban
  deployment,'' in {\em {IEEE Vehicular Technology Conference (VTC Spring)}},
  pp.~1--5, 2015.

\bibitem{Pradnya}
P.~K. Taksande, P.~Jha, and A.~Karandikar, ``Dual connectivity support in {5G}
  networks: An {SDN} based approach,'' in {\em IEEE Wireless Communications and
  Networking Conference}, pp.~1--6, 2019.

\bibitem{galeshapley}
D.~Gale and L.~S. Shapley, ``College admissions and the stability of
  marriage,'' {\em The American Mathematical Monthly}, vol.~69, no.~1,
  pp.~9--15, 1962.

\bibitem{NS3}
``{NS-3}.'' https://www.nsnam.org/.

\end{thebibliography}
\end{document}